\documentclass[review]{elsarticle}
\usepackage{float}
\usepackage{geometry}
\usepackage{epstopdf}
\usepackage{hyperref}
\usepackage{amsmath}
\usepackage{multirow}

\journal{Journal of \LaTeX\ Templates}

\begin{document}

\begin{frontmatter}

\title{Equal heartbeat intervals and their effects on the nonlinearity of permutation-based time irreversibility in heart rate}

\author[mymainaddress]{Wenpo Yao}
\author[myfirstaddress]{Wenli Yao}
\author[mythirdaddress]{Jun Wang}

\address[mymainaddress]{School of Telecommunications and Information Engineering, Nanjing University of Posts and Telecommunications, Nanjing 210003, Jiangsu, China}
\address[myfirstaddress]{School of Mines, China University of Mining and Technology, Xuzhou 221116, China}
\address[mythirdaddress]{Smart Health Big Data Analysis and Location Services Engineering Lab of Jiangsu Province, Nanjing University of Posts and Telecommunications, Nanjing 210023, Jiangsu, China}

\begin{abstract}
The association of equal heartbeat intervals with cardiac conditions and the effect of the equality on permutation-based time irreversibility are investigated in this paper. We measure the distributions of equal heartbeat intervals under three conditions, namely congestive heart failure (CHF), healthy young and elderly, whose time irreversibility of heartbeats is detected by measuring the probabilistic difference between permutations instead of raw vectors. We demonstrate that heartbeats contain high rates of equal states, particularly the CHF with around 20\% equalities, and the distributions of equal values discriminate the heartbeats at very short data length. The CHF have more equal values than the healthy young (p$<$1.47$*$$10^{-15}$) and elderly (p$<$2.48$*$$10^{-11}$), and the healthy young have less equalities than the elderly (p$<$3.16$*$$10^{-4}$). Time irreversibility considering equal values is promising to extract nonlinear behaviors of heartbeats, confirming the decreased nonlinear complexity of the diseased and aging heart rates, while that involving no equality leads to erroneous nonlinearity detection. In our contribution, we highlight the pathological or physiological information contained by the distribution of equal heartbeat intervals that might contribute to develop relevant biomarkers in the area of heart analysis, and demonstrate the effectiveness of equality-based time irreversibility in the nonlinearity detection of heartbeats.
\end{abstract}


\end{frontmatter}


\section{Introduction}
Electrocardiography (ECG) is a common process of recording the cardiac electrical activities to obtain the heart structure and/or function. Derived from the ECG, different cardiac intervals, like the PR interval (from the beginning of P wave to QRS complex), QRS interval, ST interval (between the end of QRS complex and the start of T wave), etc, carry a lot of important information, among which the RR interval between successive R waves, generally used to represent heart rate, has more clinical and scientific applications and arguably manifests with nonlinear properties \cite{Malik1996,Billman2011,Kleiger2005}. As a consequence, besides the time and frequency domains techniques, nonlinear dynamics is adopted to detect subtle nonlinear changes in heart rate to provide indices of cardiac autonomic regulation, such as Poincare plot \cite{Khandoker2013}, fractal measures \cite{Peng1995,Ivanov1999}, symbolic dynamics \cite{Kurths1995,Daw2003,Yao2017aip,Yao2018phya}, entropy methods \cite{Richman2000,Costa2002,Xiong2017}, and so on \cite{Voss2009,Shiogai2010}.

Time irreversibility is one of fundamental features to characterize nonequilibrium systems, like the nonlinear dynamical heartbeats. The statistical methodology of time irreversibility (directionality) describes a process whose probabilistic properties depend on time direction. From the statistical definitions, the quantification of time irreversibility involves in measuring the difference of joint probability distributions \cite{Weiss1975,Kelly1979}, which is not trivial, therefore some simplified alternatives are proposed. In Costa \cite{Costa2005,Costa2008} and Port \cite{Porta2006,Porta2008} parameters, temporal asymmetry is measured based on the probability divergence of ups and downs for the discrete dynamical heartbeats. L. Lacasa et al. \cite{Lacasa2012} estimate the degree of irreversibility using the Kullback¨CLeibler divergence between the in and out degree distributions based on horizontal visibility graph. And some symbolic methods, like the 'false flipped symbols' proposed by C. Daw \cite{Daw2000}, a data compression method introduced by M. Kennel \cite{Kennel2004}, the ternary coding symbolic approach provided by C. Cammarota \cite{Cammarota2007} and so forth \cite{Graff2013,Parlitz2012}, are proposed and show promising nonlinearity detection. J. Martinez et al. \cite{Martinez2018} detect time reversibility by measuring the Jensen-Shannon divergence of time forward as well as its time-reversed counterpart by means of permutation, and M. Zanin et al. \cite{Zanin2018} adopt the KL divergence to compare the probability distributions of symmetric order patterns. Considering the existence of forbidden permutation, W. Yao et al. \cite{Yao2018PLA} propose a subtraction-based parameter to measure the probabilistic difference between order patterns for the time irreversibility. These simplified approaches have been gaining growing popularity for the features of fast, simplicity, noise insensitivity and so on.

Regarding the discrete heartbeats, HRV (heart rate variability) is defined as the physiological phenomenon of variation in successive heartbeats. To understand the physiological basis that underlies HRV, intensive investigations have been conducted. Variability in heartbeats is subject to the sympathetic and the parasympathetic nervous system, and the reduced HRV, extensively supported by experimental and clinical reports \cite{Malik1996,Billman2011,Wolf1978,Tsuji1994,Huikuri2001,Thayer2010}, is a promising indication of increased risk for severe ventricular arrhythmia and sudden cardiac mortality. The low variability between heartbeats might lead to equal RR intervals under low precision of data collection or R wave detection \cite{Bian2012}, therefore the equal states may contain important meaningful information about cardiac autonomic modulation. However, equal states and their effects on nonlinear dynamics analysis in heartbeats are not given deserved consideration.

In the Costa \cite{Costa2005,Costa2008} and Port \cite{Porta2006,Porta2008} parameters, considering only two neighboring heartbeats, equalities imply reversibility which is not reasonable if we consider more than two values. Equal values in the original permutation entropy \cite{Bandt2002} are neglected considering the continuous distribution of time series and rare equal values, which is true in some real-world signals like the brain electrical activities \cite{Yao2018PLA} while is not rational in heartbeats analysis. In the implementation of permutation, equal values may also introduce a significant spurious effect in practical contexts and lead to erroneous conclusions \cite{Zunino2017}. C. Bian et al. \cite{Bian2012} report that equal states play important role in permutation entropy analysis of heartbeats and propose a modified equality-involved permutation that allows more accurate characterization of heartbeat dynamics.

In our contribution, we conduct research on the association of equal RR intervals with cardiac physiological or pathological conditions, and analyze the effects of equality on the heartbeats' time irreversibility quantified by the probabilistic differences between permutations. Three groups of heart rates, of congestive heart failure (CHF), healthy young and elderly subjects, from the public PhysioNet \cite{Goldberger2000} are collected for our research.

\section{Methods}

\subsection{Time irreversibility}
\begin{figure}[htb]
  \centering
    \includegraphics[width=8.3cm,height=4.5cm]{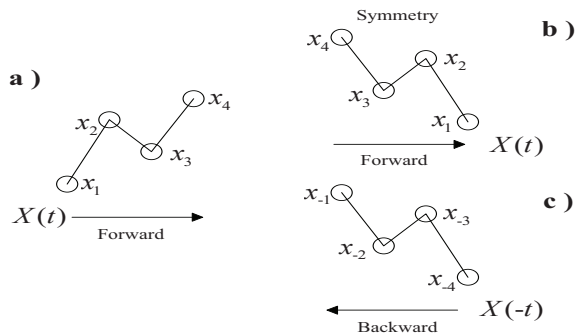}
  \caption{Exemplary vectors in forward and backward time series and the symmetric vector. For the vector of $(x_{1},x_{2},x_{3},x_{4})$ in a), its symmetric one $(x_{4},x_{3},x_{2},x_{1})$ in b) and the corresponding form $(x_{-1},x_{-2},x_{-3},x_{-4})$ in c) in reverse series are same.}
  \label{fig1}
\end{figure}

Time reversibility, or the term of temporal asymmetry, is a property being defined that time series is invariant with respect to time reversal \cite{Weiss1975,Kelly1979}, and reversibility is a property that is not affected by arbitrary static transformation \cite{Ramsey1995}. Statistically speaking, if a process $X(t)$ is time reversible, vector $\{X(t_{1}),X(t_{2}),\cdots,X(t_{m})\}$ and $\{X(-t_{1}),X(-t_{2}),\cdots,X(-t_{m})\}$ in reverse series and its symmetric form $\{X(t_{m}),\cdots,X(t_{2}),X(t_{1})\}$ for every $t$ and $m$ have the same joint probability distributions. It is equivalent to measure the temporal asymmetry based on the probabilistic differences between symmetric joint distributions and quantify the time irreversibility based on the probabilistic difference between forward and backward time series, demonstrated visually in Fig.~\ref{fig1}.

It is difficult to calculate the joint probability distributions of time series, and to simplify the quantification of time irreversibility, several approaches have been proposed \cite{Costa2005,Costa2008,Porta2006,Lacasa2012,Daw2000}. Among these simplified alternatives, methods based on order patterns have gained growing popularity for its simplicity and close connection with time irreversibility \cite{Martinez2018,Zanin2018,Yao2018PLA}.

\subsection{Time irreversibility based on permutation}
The permutation method, coming naturally from time series without further model assumptions, is originally introduced by C. Bandt and B. Pompe in the permutation entropy \cite{Bandt2002}, a complexity parameter based on comparison of neighboring values. Since then, the permutation entropy and the order pattern scheme have attracted much attention with a huge number of applications \cite{Bandt2016}, and some modifications or improvements are proposed, such as non-uniform embedding \cite{Tao2018}, equality-involved and weighted permutation entropy \cite{Bian2012,Fadlallah2013}, etc.

\begin{figure}[htb]
  \centering
    \includegraphics[width=8.3cm,height=3.7cm]{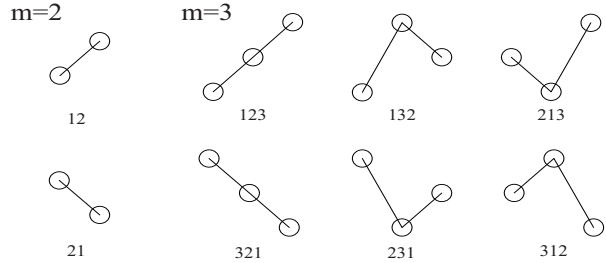}
  \caption{Order patterns without equal values when $m$=2 and 3.}
  \label{fig2}
\end{figure}

Let us recall the basic permutation method. Given time series $X(t)=\{x_{1},x_{2},\cdots,x_{t}\}$, we construct embedding space $X_{m}^{\tau}(i)=\{x(i),x(i+\tau),\ldots,x(i+(m-1)\tau)\}$ for dimension $m$ and delay $\tau$. And then, we reorder the elements in each space vector according to their relative values, for example in ascending order, $x(j_{1}) \leq x(j_{2}) \leq \cdots \leq x(j_{i})$, and map the vector onto order pattern $\pi_{j}=(j_{1},j_{2}, \cdots, j_{i})$. Fig.~\ref{fig2} displays order patterns of $m$=2 and 3 whose upper bound are 2=2! and 6=3!.

To simplify the quantification of time irreversibility, symmetric vectors $X_{j}=\{x_{1},x_{2},\cdots,x_{m}\}$ and $X_{s}=\{x_{m},\cdots,x_{2},x_{1}\}$ could be replaced by their order patterns $\pi_{j}$ and $\pi_{s}$ whose probability distributions are $p(\pi_{j})$ and $p(\pi_{s})$. Concerning the existence of forbidden permutations that some order patterns may not have counterparts, the present authors \cite{Yao2018PLA} proposed a subtraction-based parameter $Y_{s}$, in Eq.~\ref{eq1}, to measure the probabilistic divergence between $p(\pi_{j})$ and $p(\pi_{s})$.

\begin{eqnarray}
\label{eq1}
Y_{s} = \sum  p(\pi_{j})\frac{p(\pi_{j})-p(\pi_{s})}{p(\pi_{j})+p(\pi_{s})}
\end{eqnarray}

In the original permutation \cite{Bandt2002}, equal states are neglected under a weak stationarity assumption that time series have continuous distributions and equal states are very rare. The rare equalities are broken by adding random perturbations or treated as ups or downs according to their orders of occurrence in several areas \cite{Martinez2018,Zanin2018,Yao2018PLA,Bandt2016,Yao2014}. The neglect of equal values is reasonable in some applications, such as the brain electrical activities with rare equalities, however, in other situations, where equalities are not rare or the equal values contain important underlying information about systems \cite{Bian2012,Zunino2017}, it is not rational. For example, equal heartbeat intervals are highly relevant to physiological or pathological conditions and have effects on heart nonlinear dynamics analysis \cite{Bian2012}.

\subsection{Equality-involved permutation}
Taking the equal values into consideration, C. Bian and Q. Ma \cite{Bian2012} proposed a modified permutation method by replacing the successive indexes of equal values to the smallest ones to improve the permutation entropy in heartbeat analysis.

In the original permutation, equal values, for example $x(j_{i})=x(j_{i+1})$ and $x(j_{l})=x(j_{l+1})=x(j_{l+2})$, are in adjacent continuous orders if we organize them according to their orders of occurrence, and the indexes of the equal states in permutation are successive $(j_{i},j_{i+1})$ and $(j_{l},j_{l+1},j_{l+2})$. The modified method is to revise the values in each set of successive indexes to be identical to the smallest one in their related index sets, then $(j_{i},j_{i+1})$ and $(j_{l},j_{l+1},j_{l+2})$ will be rewritten to $(j_{i},j_{i})$ and $(j_{l},j_{l},j_{l})$. Taking vector $\{2,2,1,3,1,2\}$ as an example, its ascending reorganization is $\{1,1,2,2,2,3\}$, and the order pattern in the original permutation method is (351264) while in the equality-involved approach the order pattern should be modified to (331114).

\begin{figure}[htb]
  \centering
    \includegraphics[width=6.3cm,height=3.9cm]{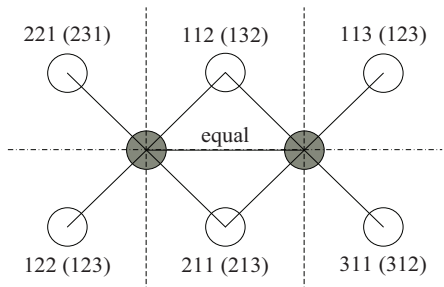}
  \caption{Modified permutations and original order patterns (in parentheses) of $m$=3 when there are two equal values (in gray).}
  \label{fig3}
\end{figure}

In the modified permutation, to determine the upper bounds is more complicate than the original $m$!. When $m$ is 2, there are three order patterns, up, down and equality, and when $m$ is 3, besides the 6 permutations in Fig.~\ref{fig2}, there are 6 more order patterns considering two equal values, illustrated in Fig.~\ref{fig3}, and a triple-equal order type of '111'. C. Bian  and Q. Ma provided a recursive method in their contribution together with the upper bounds (13, 73, 501, 4051 and 37633) of $m$ from 3 to 7.

\section{Equal values in heart rates}

\subsection{Heart rates from the public PhysioNet}
\begin{figure}[htb]
  \centering
    \includegraphics[width=6.3cm,height=3.9cm]{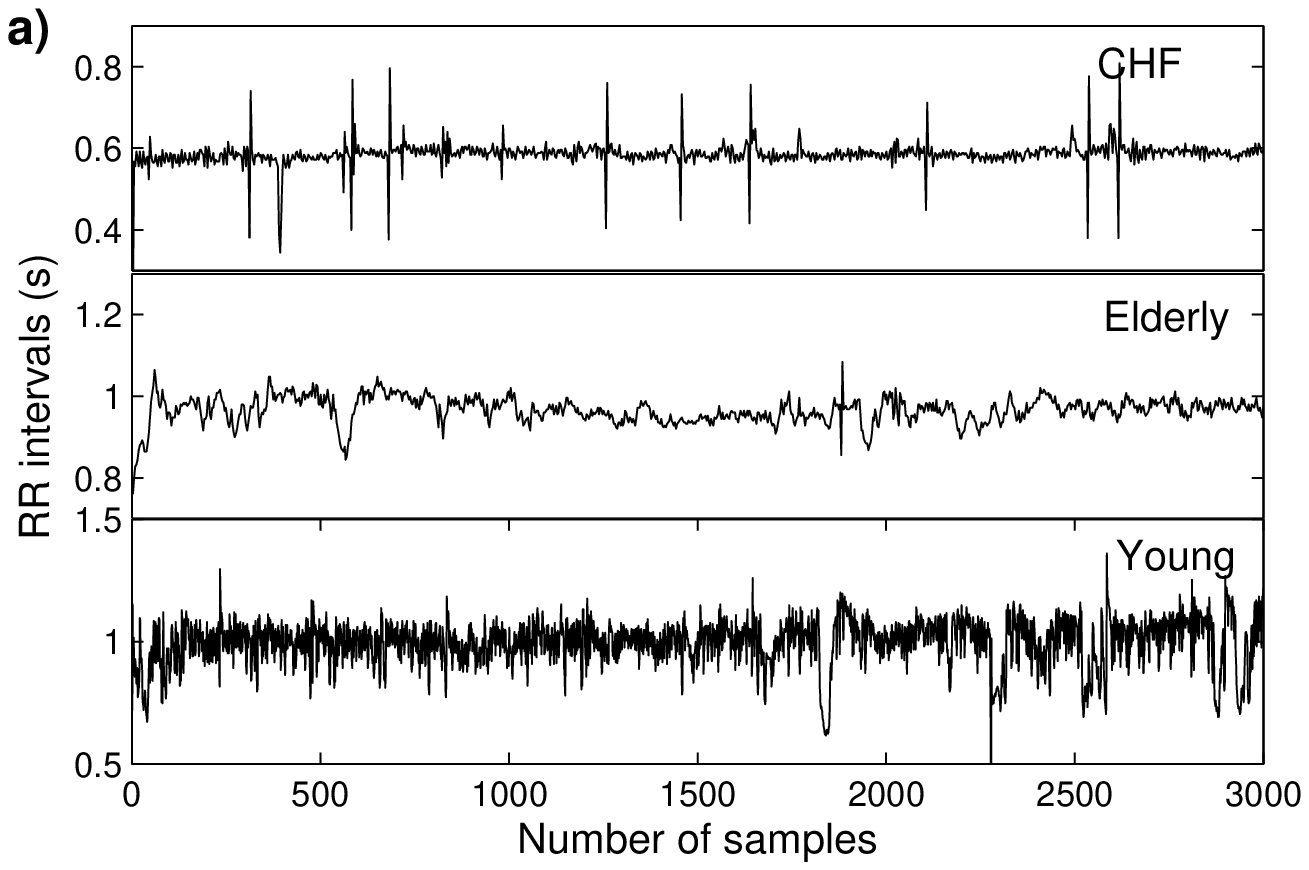}
    \includegraphics[width=6.3cm,height=3.9cm]{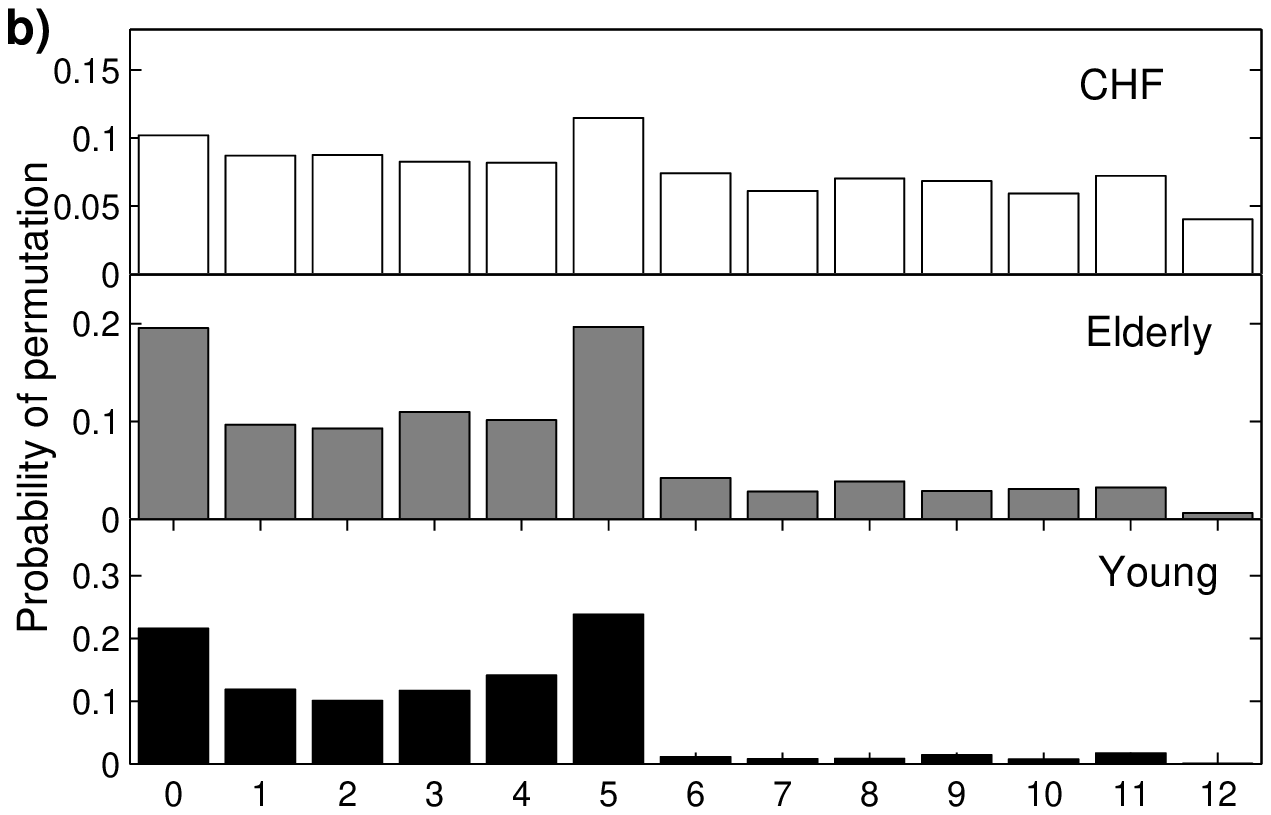}
  \caption{Heartbeats and the equality-involved permutations. Of the subplot b), 0 to 5 in x-label denote permutation without equal values, 6 to 11 refer double-equal order patterns and 12 represents the triple-equal '111'.}
  \label{fig4}
\end{figure}

Three groups of heart data from PhysioNet \cite{Goldberger2000} are collected in our study. Two groups of healthy subjects, 20 young
(mean age 25.8$\pm $4.3 years, range 21 to 34 years) and 20 elderly (mean age 74.5$\pm $4.4 years, range 68 to 85 years) volunteers, of the Fantasia database \cite{Iyengar1996} contribute 120 minutes of continuous ECG collection. The CHF data sets \cite{Baim1986}
include long-term ECG recordings from 44 patients (mean age 55.5$ \pm $11.4 years, range 22 to 79 years), of which 15 subjects (11 men, aged 22 to 71, and 4 women, aged 54 to 63) are from the `chfdb' database and 29 subjects (aged 34 to 79) are from the `chf2db' database. Sampling frequency of the ECG in the `fantasia' and `chfdb' databases is 250 Hz and those in the `chf2db' is 128 Hz, and the ADC (analog to digital converters) resolution is 16 bit (the f1* records in `fantasia' database) or 12 bit (the records in `chfdb' and f2* records in `fantasia' database). Each set of heartbeat derived from ECG is manually reviewed and corrected by data providers or the present authors. Exemplary heart rates and the distributions of equality-involved permutations of $m$=3 are illustrated in in Fig.~\ref{fig4}.

In Fig.~\ref{fig4}a, CHF heart rates have low variability and low amplitude resolutions while the healthy young has high variability. In Fig.~\ref{fig4}b, the equality-involved permutations of the CHF heartbeats have comparative distribution probabilities to those without equal values, all around 6\% to 8\%, and the probability of triple equality '111' represented by '12' is about 5\%. Permutations involving equal values in the two groups of healthy heartbeats, especially the healthy young, are much less than those without equalities.

Equal values exist in heartbeats and their distribution probabilities are different under different cardiac conditions, suggesting the equal heartbeat intervals might have connections with cardiac physiological and pathological conditions.

\subsection{The distribution of equal states in heart rates}
Given neighboring equal state, $x_{i}=x_{i+\tau}$, $\tau \neq 0$, we define the rate of equal values of time series as Eq.~\ref{eq2}, where $N(x_{i}=x_{i+\tau})$ is the amount of equal values and $N(x_{i})$ is the total number of heart beats.
\begin{eqnarray}
\label{eq2}
eR = \frac{N(x_{i}=x_{i+\tau})}{N(x_{i})-1}
\end{eqnarray}

\begin{figure}[htb]
  \centering
    \includegraphics[width=7.2cm,height=4.5cm]{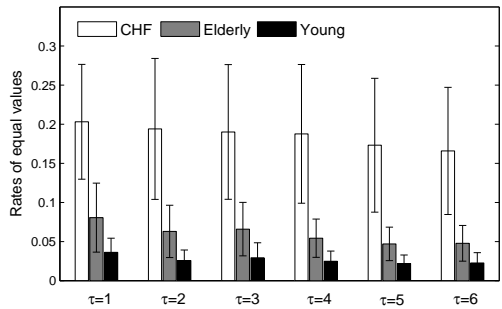}
  \caption{Rates of neighboring equal values (mean$\pm$std) in CHF, healthy elderly and young heartbeats.}
  \label{fig5}
\end{figure}

Let us look at the distributions of equal states in the CHF, healthy elderly and young heart rates, shown in Fig.~\ref{fig5}.

From Fig.~\ref{fig5}, equalities in the CHF heartbeats are much more than those of the two healthy groups, and $eR$ of the healthy young are the least. In the CHF heart beats, the mean rates of neighboring equalities are close to 20\%. The distribution of equal values of the healthy young is the lowest 2.6\%, and equal values in the healthy elderly account about 6.5\%, in-between the CHF and the healthy young. And we find the highest equality rate of 46.11\% is from the set of heartbeats identified as 'chf226' in the 'chf2db' when $\tau$=2. Statistical tests (p value of independent sample t test) for $eR$ of the three groups of heart rates are listed in Table~\ref{tab1}.

\begin{table}[htb]
\centering
\caption{Independent sample t test of the heartbeats' rates of equal values. 'C', 'E' and 'Y' represent the CHF, healthy elderly and young heartbeats.}
\label{tab1}
\begin{tabular}{c|ccc}
\hline
$\tau$ &C-E	&C-Y	 &E-Y \\
\hline
1 &2.48*$10^{-11}$ &1.10*$10^{-19}$ &3.16*$10^{-4}$ \\
2 &7.78*$10^{-12}$ &4.75*$10^{-16}$ &9.90*$10^{-5}$ \\
3 &1.61*$10^{-11}$ &3.26*$10^{-16}$ &2.29*$10^{-4}$ \\
4 &8.76*$10^{-13}$ &8.76*$10^{-16}$ &5.10*$10^{-5}$ \\
5 &1.47*$10^{-12}$ &1.47*$10^{-15}$ &6.80*$10^{-5}$ \\
6 &2.81*$10^{-12}$ &2.81*$10^{-16}$ &1.87*$10^{-4}$ \\
\hline
\end{tabular}
\end{table}

The rates of neighboring equal values in heart rates significantly discriminate the CHF, healthy young and elderly subjects, and the differences between each two groups¡¯ distribution of double-equal values are not affected by time delay. When $\tau$=4, the CHF-elderly (p=8.76*$10^{-13}$) and elderly-young (p=5.10*$10^{-5}$) heartbeats have best discrimination and the best difference of the CHF-young (p=1.10*$10^{-19}$) group comes when $\tau$=1. CHF-elderly p values are generally smaller than 3.0*$10^{-11}$, and the CHF-young p values are not bigger than 4.0*$10^{-15}$, the difference of between the two groups healthy subjects (p$<$0.0004) although not that satisfied still significant and statistically acceptable, suggesting the rates of equal values could represent a promising parameter for differentiating heart conditions.

\begin{figure}[htb]
  \centering
    \includegraphics[width=5.9cm,height=4.5cm]{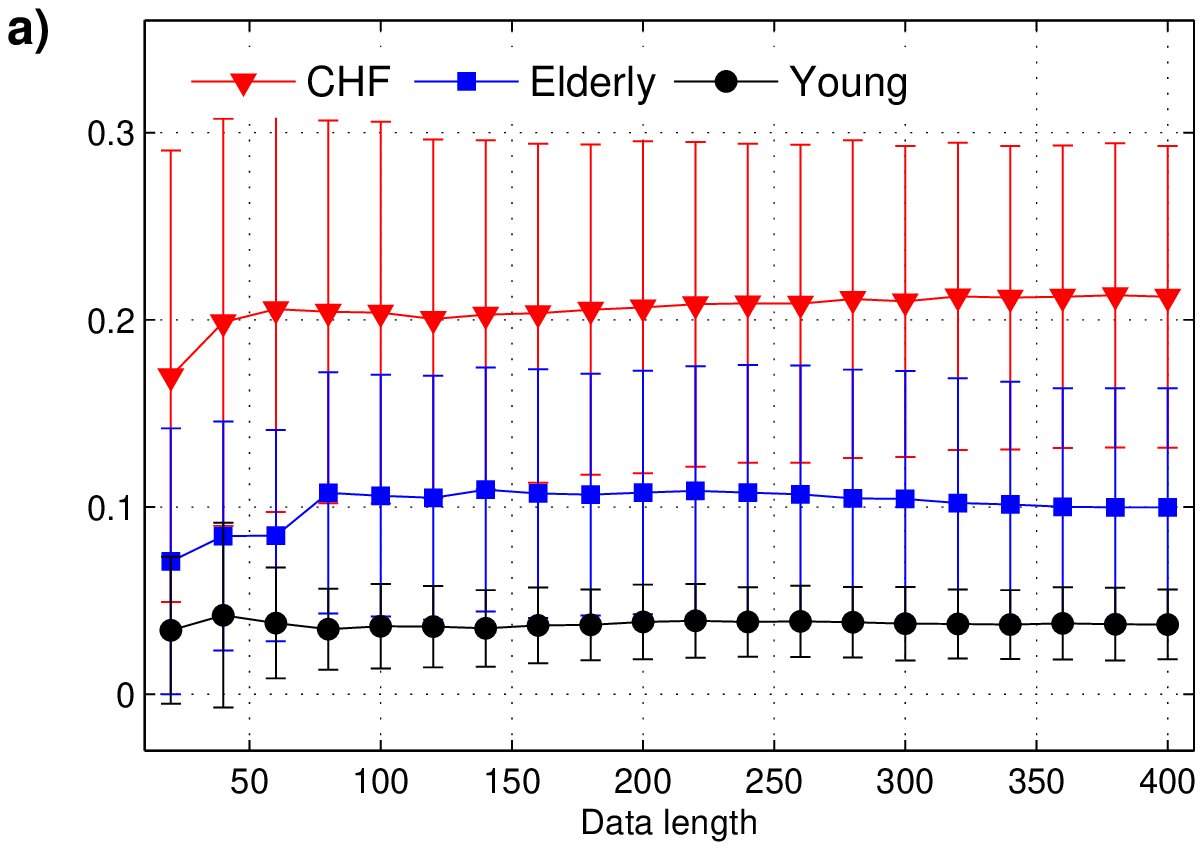}
    \includegraphics[width=5.9cm,height=4.5cm]{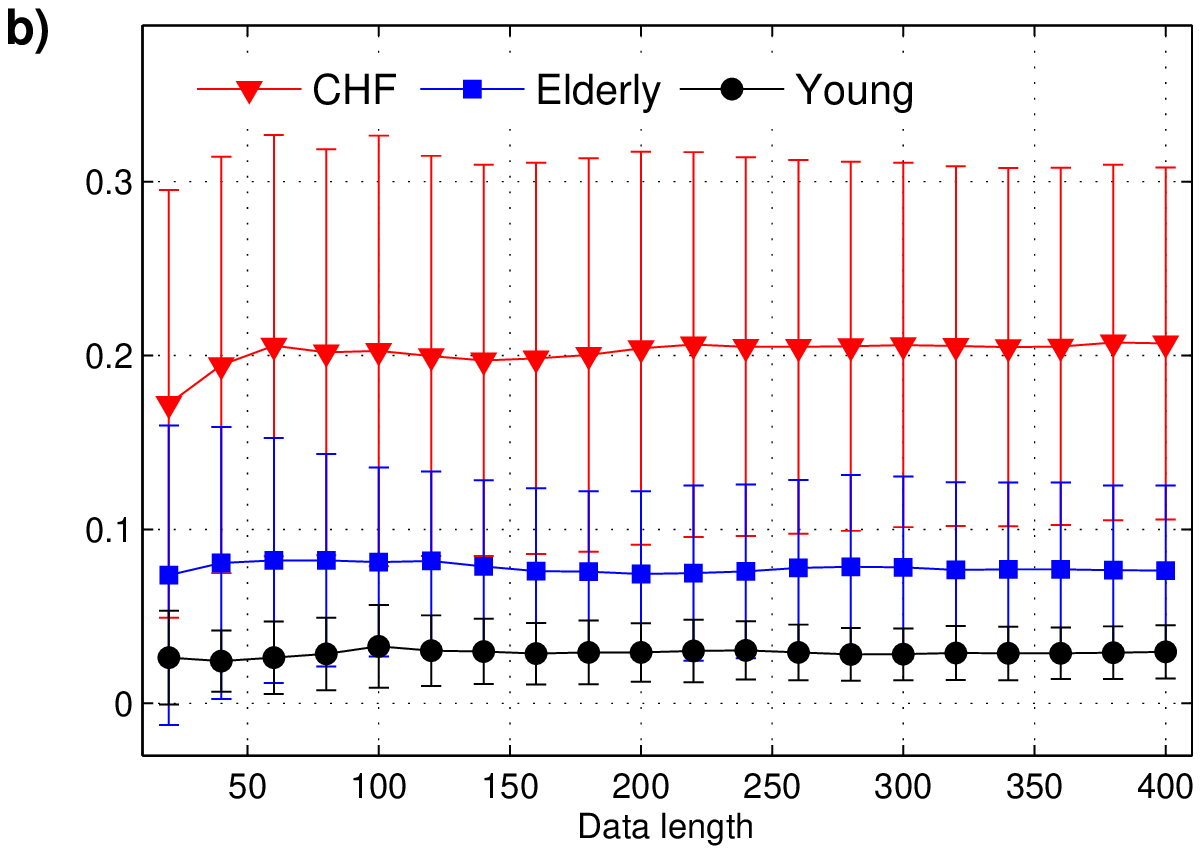}
  \caption{Rates of equal RR intervals (mean$\pm$std) in CHF, healthy heartbeats with changing data length. a) $\tau$=1. b) $\tau$=2.}
  \label{fig5+}
\end{figure}

Moreover, the rates of equal RR intervals differentiate the three groups of heart rates at very short data length, illustrated in Fig.~\ref{fig5+}.

When data length is bigger than 80, $eR$ of the three groups of heartbeats become convergent and the statistical discriminations are all acceptable (p$<$0.0001). And we find that the starting points have no significant influence to the results, suggesting the distribution of equal values in heartbeats could reliably characterize the three cardiac conditions.

The connection between slower heart rates and acute myocardial infarction is first demonstrated by Wolf et al. \cite{Wolf1978}, which has been subsequently confirmed by numerous representative clinical reports and following investigators \cite{Malik1996,Billman2011,Tsuji1994,Thayer2010}. Specifically, patients recovering from myocardial infarctions have reduced variability in heart rates, further, those with greater low HRV also have greater risk for sudden death. Investigators have demonstrated that patients with myocardial infarctions have the greatest risk of sudden death if they have smallest HRV. Reduced HRV now has been serve as a strong independent predictor of mortality following infraction due to structural heart changes, although the exact biological mechanism accounting for the association of reduced HRV with mortality is unknown. Reduced variability in heartbeats leads to increased possibility of equal R-R intervals, and therefore the increased rate of equal heartbeats interval contains valuable information about cardiac regulation.

As the reduced HRV is a clinical indicator to sudden death, the increased distribution of equal values in heart rates, quantifying the degree of reduction of variability in heart rates, might serve as an independent indicator for some cardiac physiological or pathological conditions clinically.

\section{Time irreversibility of the heart rates}
The discrete dynamical heartbeats manifest the property of time irreversibility \cite{Costa2005,Costa2008,Porta2006,Porta2008}. In this section, we employ $Y_{s}$ to measure the probabilistic difference of permutations instead of raw vectors for time irreversibility, and analyze the effects of equal states on the nonlinearity detection of heart rates.

We generate 100 surrogate data sets for each set of heartbeat to testify the nonlinearity by determining whether time irreversibility of the original data is significantly different from the surrogate \cite{Theiler1992}, say, smaller than the 2.5th percentile or bigger than 97.5th percentile of the surrogate data set. The improved amplitude adjusted Fourier transform (iAAFT) \cite{Schreiber1996} that consists of iterative random shuffle and Fourier transform of time series is adopted in this paper. The iAAFT generates surrogate data with same autocorrelations, probability distribution and power spectrum to the original data and keeps the linear correlations and destroy all the nonlinear ones.

\subsection{$Y_{s}$ of $m$=2 and Costa parameter}
Let us first use the Costa parameter to detect the nonlinearity of temporal asymmetry. Defining the difference between neighboring heartbeats as $\Delta x=x(i+\tau)-x(i)$, $\tau \neq 0$, of which $\Delta x^{+}$ denotes up and $\Delta x^{-}$ represents down, Costa \cite{Costa2005} proposed $\widehat{A}=\frac{\sum Prln(-\Delta x^{-})-\sum Prln(\Delta x^{+})}{\sum Prln(\Delta x)}$ for temporal asymmetry of heartbeats, and then simplified the parameter to $A=\frac{\sum H(-\Delta x^{-})-\sum H(\Delta x^{+})}{N(\Delta x )-1}$ \cite{Costa2008} that yields comparable results and is easier to implement.

Costa indexes of the CHF, healthy elderly and young heartbeats as well as their surrogate data are listed in Table~\ref{tab2}. For statistical convenience, we provide Costa parameter (mean$\pm$std) of all the surrogate data for each groups of heartbeats in Table~\ref{tab2}. We should note that Costa index of each set of heartbeat is larger than 97.5th percentile of its surrogate data sets, validating the nonlinearity of heartbeats, which is also true in the following sections.

\begin{table}[htb]
\centering
\caption{Costa parameter (mean$\pm$std) of the three groups of heartbeats and the surrogate. Costa-S denotes the Costa index of the surrogate data.}
\label{tab2}
\begin{tabular}{cccc}
\hline
 &CHF &Elderly	 &Young \\
\hline
Costa   &0.0227$\pm$0.0217	  &0.0341$\pm$0.0274	&0.0416$\pm$0.0294 \\
Costa-S   &0.0013$\pm$0.0010	&0.0053$\pm$0.0040 &0.0058$\pm$0.0044 \\
\hline
\end{tabular}
\end{table}

The healthy young have the highest time irreversibility while the CHF have the lowest, being consistent with the complexity losing theory \cite{Ivanov1999,Yao2017aip,Costa2002,Costa2005,Goldberger2002,Glass2001} of diseased and aging heartbeats. The theory of complexity loss in aging and disease lies in the reduced cardiac adaptive capabilities of individuals that CHF patients have damaged cardiac regulation, and the aging subjects have abnormalities in cardiac functionalities, while the healthy physiological systems reveal complex variability long-range correlations and distinct nonlinear interactions. Independent sample t tests for the heartbeats¡¯ time irreversibility suggest that Costa has acceptable discriminations between the CHF-young (p=0.011) group while does not discriminate the CHF-elderly (p=0.073) and elderly-young (p=0.499) groups statistically.

\begin{figure}[htb]
  \centering
    \includegraphics[width=5.9cm,height=4.5cm]{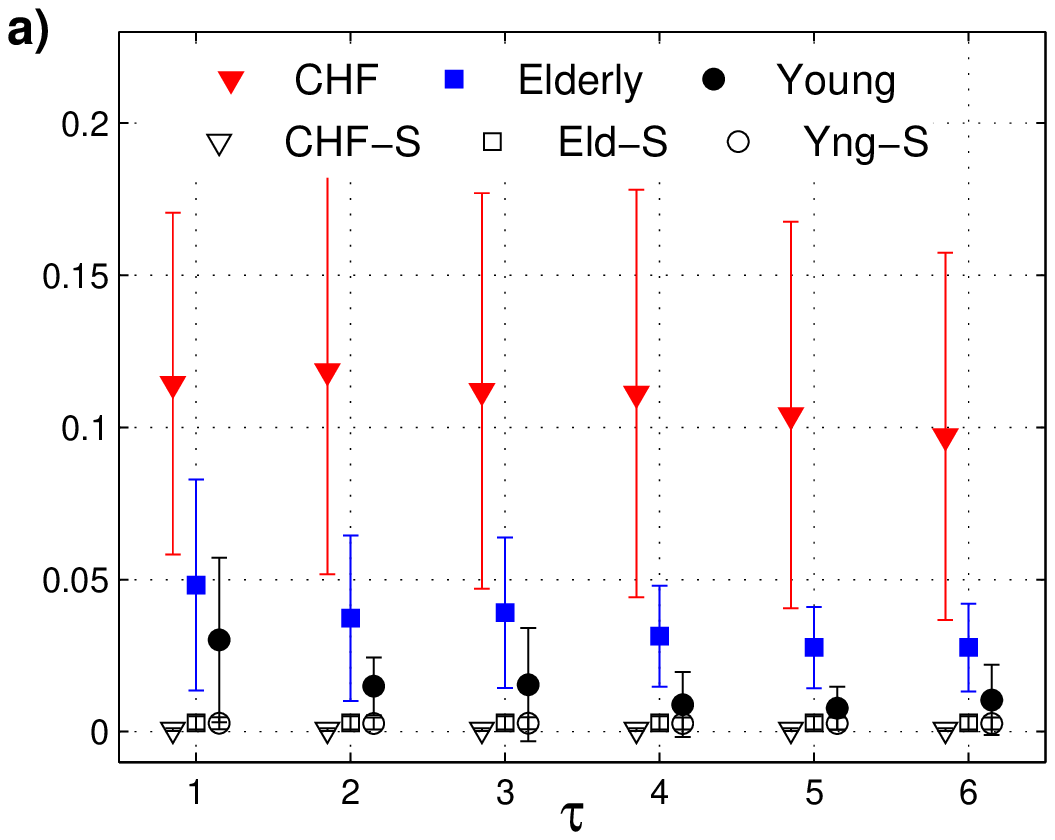}
    \includegraphics[width=5.9cm,height=4.5cm]{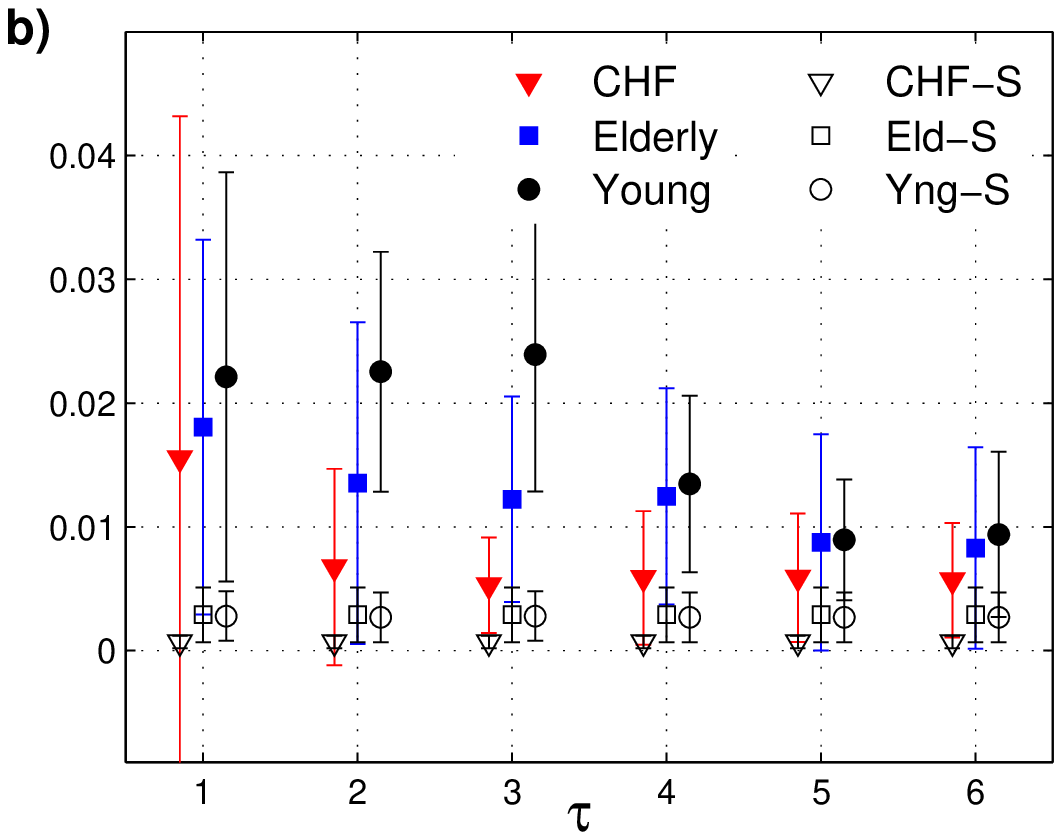}
  \caption{Time irreversibility (mean$\pm$std) of heartbeats and surrogate when $m$=2. a) $Y_{s}$ of original permutations. b) $Y_{s}$ of equality-based permutations. 'CHF-S', 'Eld-S' and 'Yng-S' represent the surrogate data of the CHF, healthy elderly and young heartbeats.}
  \label{fig6}
\end{figure}

$Y_{s}$ of probabilistic difference of symmetric permutation with and without equal values when $m$=2 are shown in Fig.~\ref{fig6}. In Fig.~\ref{fig6}a, time irreversibility based on original permutation of the three groups of heartbeats are completely contradictory to the complexity losing theory about the diseased and aging heart rates. $Y_{s}$ of the CHF heartbeats are the biggest while the healthy young have the lowest time irreversibility. In Fig.~\ref{fig6}b, considering equal states and applying the modified permutation, time irreversibility of the three kinds of heart data are rational, the healthy young $>$ the healthy elderly $>$ the CHF, in line with complexity losing theory.

When $m$=2 and $\tau$=1, if equal values are very rare, $Y_{s}$ of permutation is equivalent to the Costa parameter \cite{Yao2018PLA}, and if equalities can not be neglected, order pattern of equal states is the self-symmetric '11', then $Y_{s}$ of modified permutation, measuring the difference between ups and downs, is in line with the Costa index. $Y_{s}$ of modified permutation confirms the complexity losing theory in heartbeats while the discrimination between each two groups are not statistically acceptable (p$>$0.05), indicated by Table~\ref{tab3}.

\begin{table}[htb]
\centering
\caption{Independent sample t test of the heartbeats¡¯ time irreversibility of $Y_{s}$ when $m$=2.}
\label{tab3}
\begin{tabular}{ccccc cc}
\hline
$\tau$  &1	&2	 &3	 &4	 &5	 &6 \\
\hline
C-E &0.70640	 &0.040802	 &0.00165	 &0.00458	 &0.11587	 &0.11587 \\
E-Y &0.42238	 &0.017727	 &0.00054	 &0.69059	 &0.92693	 &0.64279 \\
C-Y &0.33137	 &6.39E-09	 &3.06E-07	 &0.00021	 &0.03152	 &0.01403 \\
\hline
\end{tabular}
\end{table}

When time delay increases to 2 or 3, $Y_{s}$ of the different heartbeats show satisfied statistical discriminations. The best discrimination between nonlinearity of the CHF and young heart rates lie in $\tau$=3, and those for the healthy elderly and other two groups is $\tau$=2. As $\tau$ increases to 4 or bigger, the heartbeats¡¯ nonlinearity although conforms the complexity losing theory, the discrimination between them deteriorate, especially for the healthy elderly and other two groups of heartbeats (p$>$0.05).

In this subsection, we note that equal values have significant effects on the time irreversibility based on permutation. $Y_{s}$ for probabilistic difference of permutation has reliable heartbeats nonlinearity detection if we take equal states into account while do not have acceptable results if we neglect the equalities. And we learn that the delay plays important role and a proper delay leads to better nonlinearity extraction.

\subsection{Equality-involved time irreversibility when $m>$2}
\begin{figure}[htb]
  \centering
    \includegraphics[width=4cm,height=4cm]{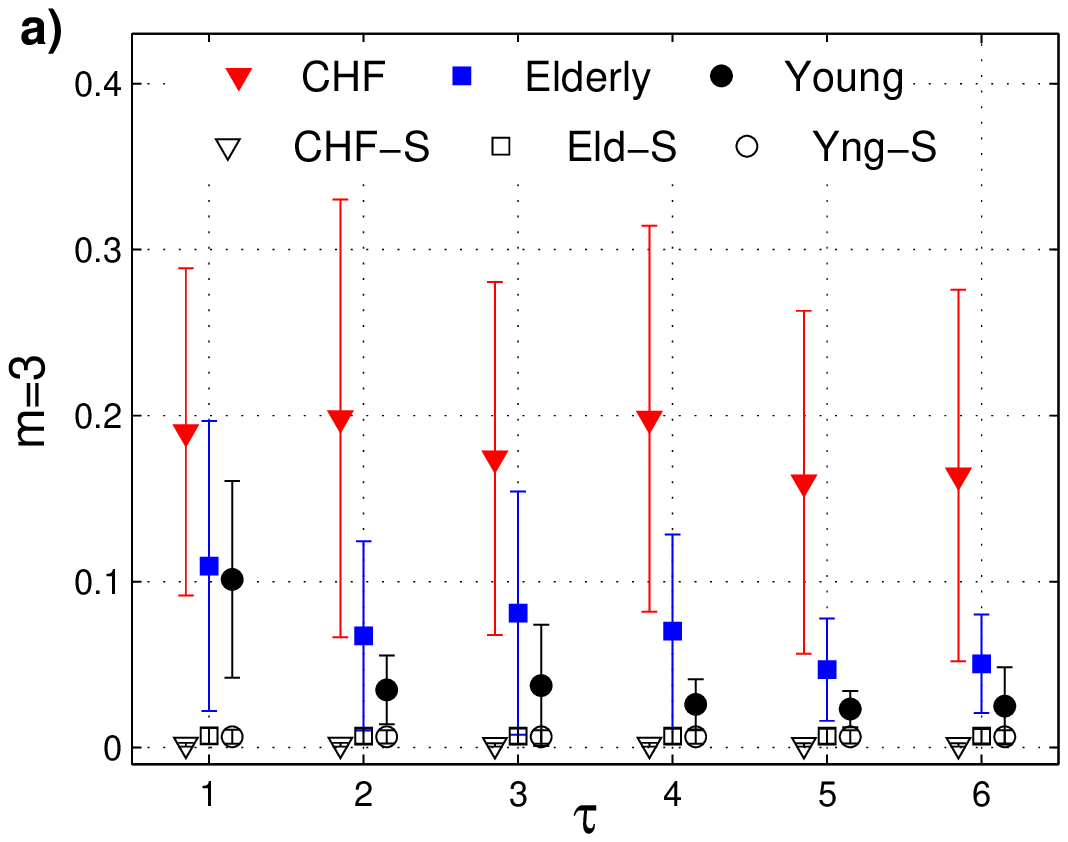}
    \includegraphics[width=4cm,height=4cm]{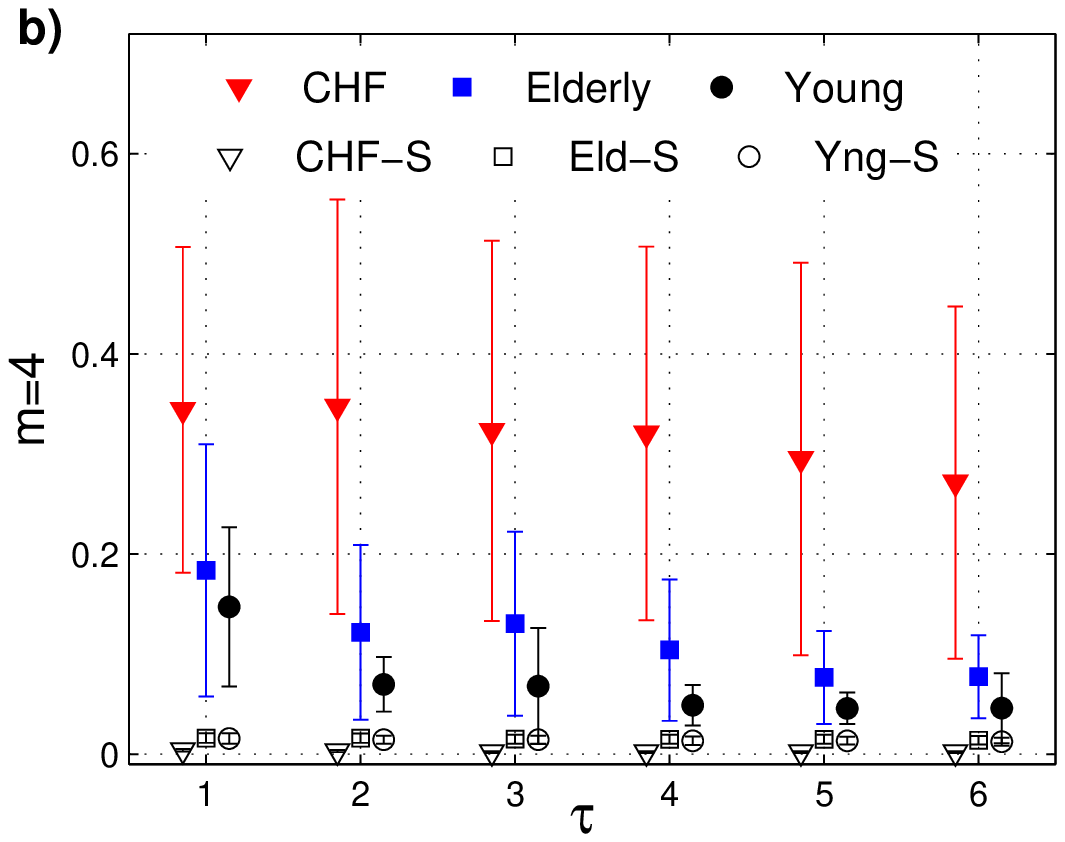}
    \includegraphics[width=4cm,height=4cm]{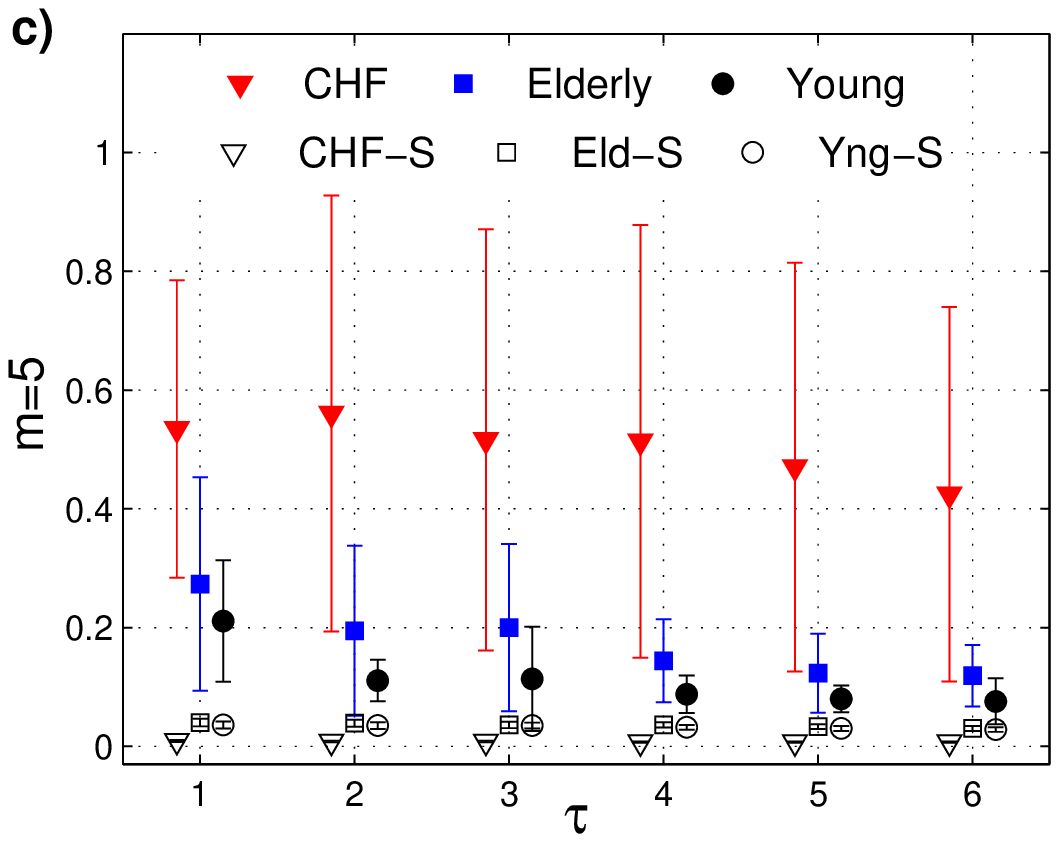}
    \includegraphics[width=4cm,height=4cm]{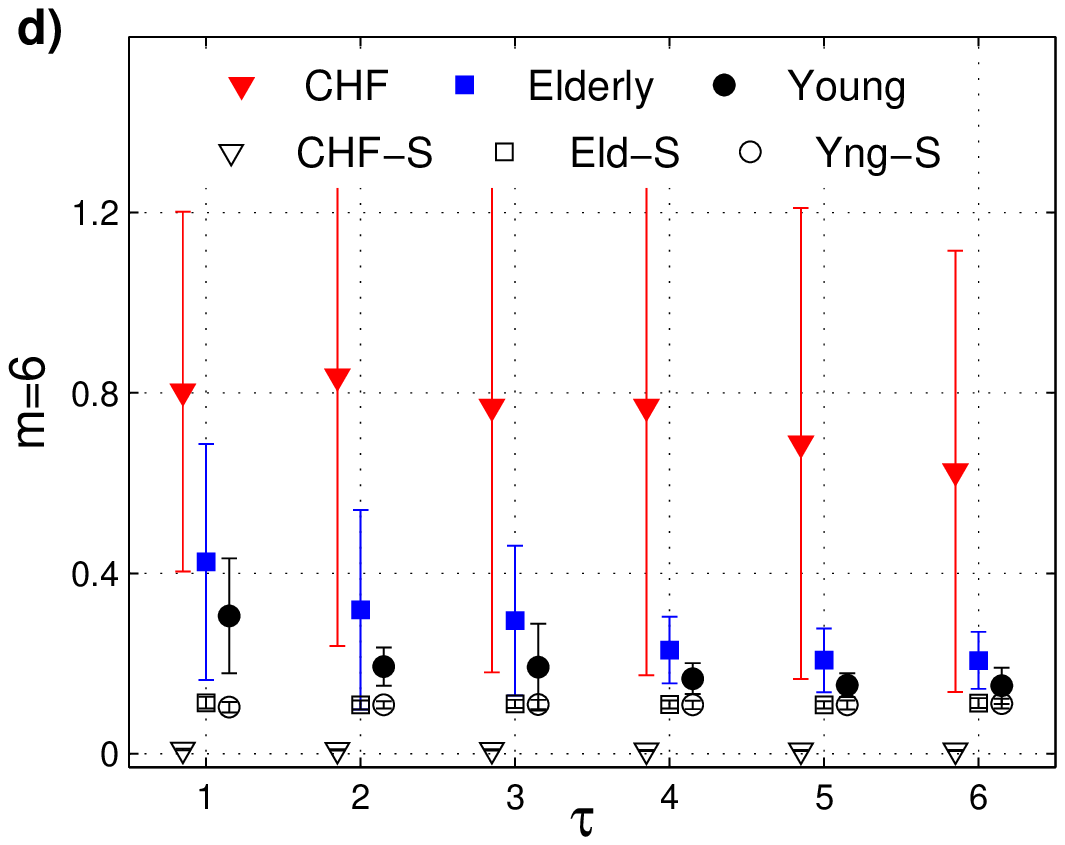}
    \includegraphics[width=4cm,height=4cm]{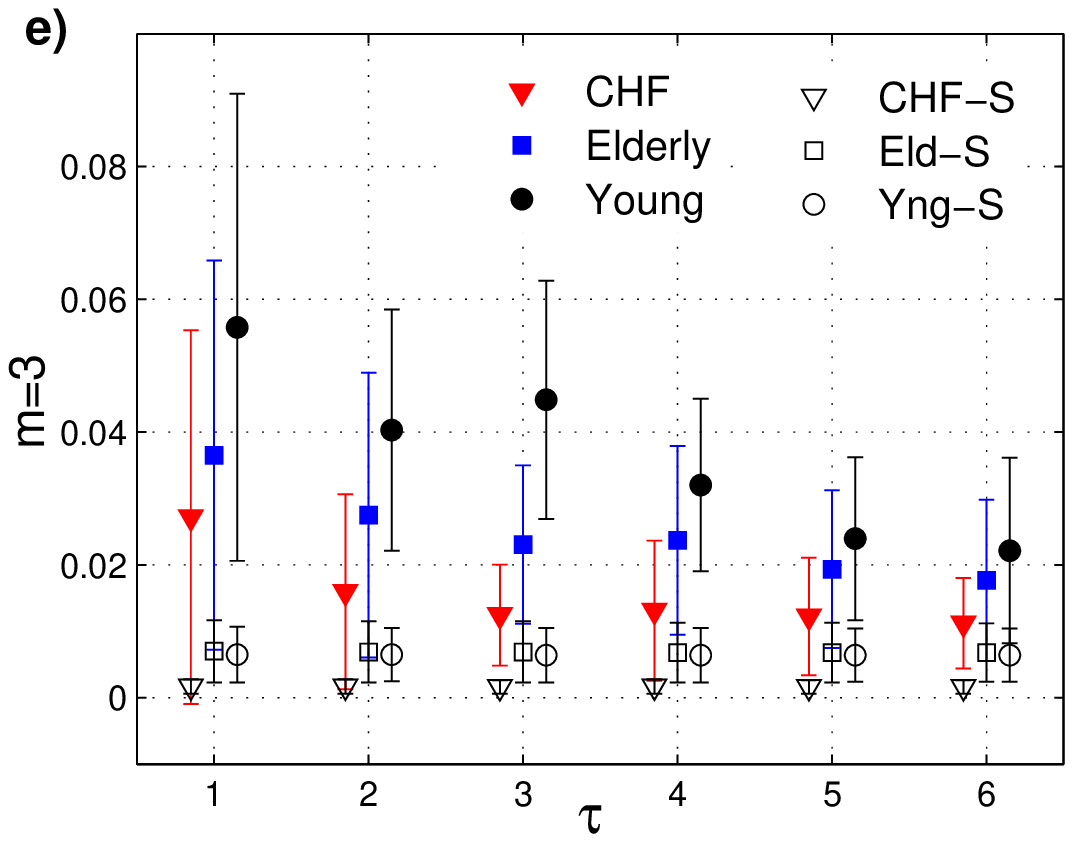}
    \includegraphics[width=4cm,height=4cm]{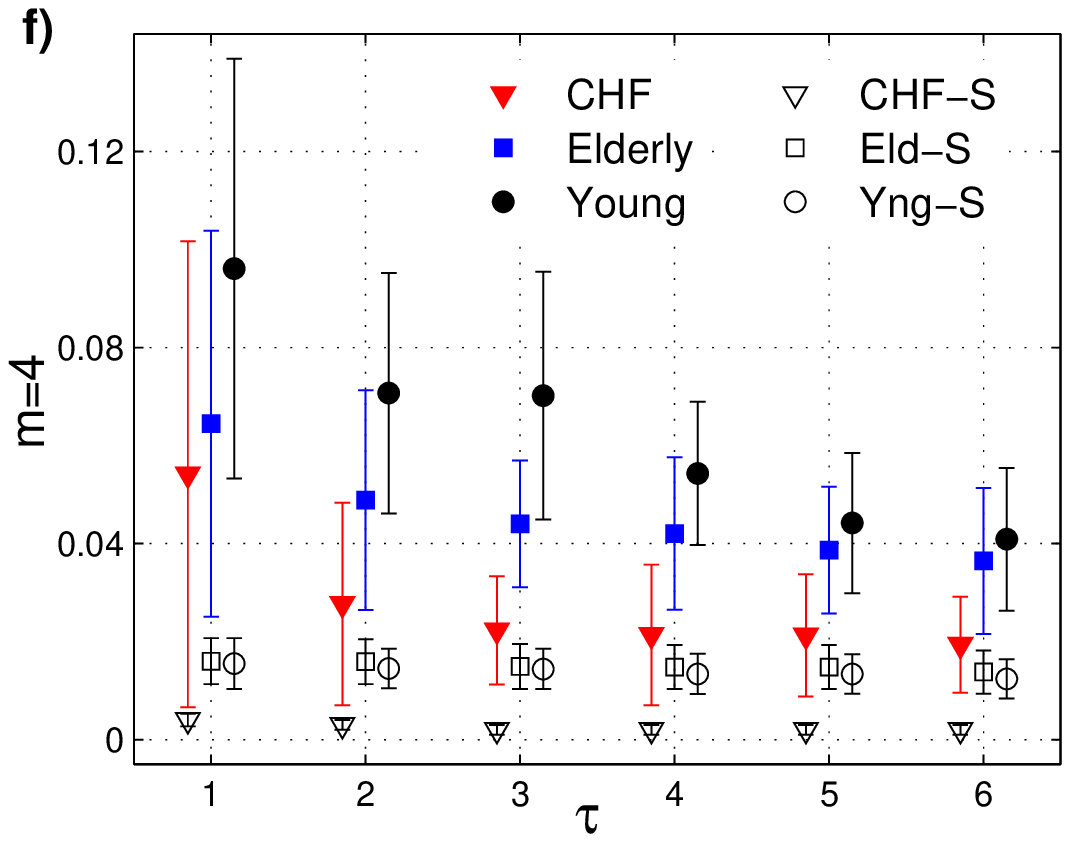}
    \includegraphics[width=4cm,height=4cm]{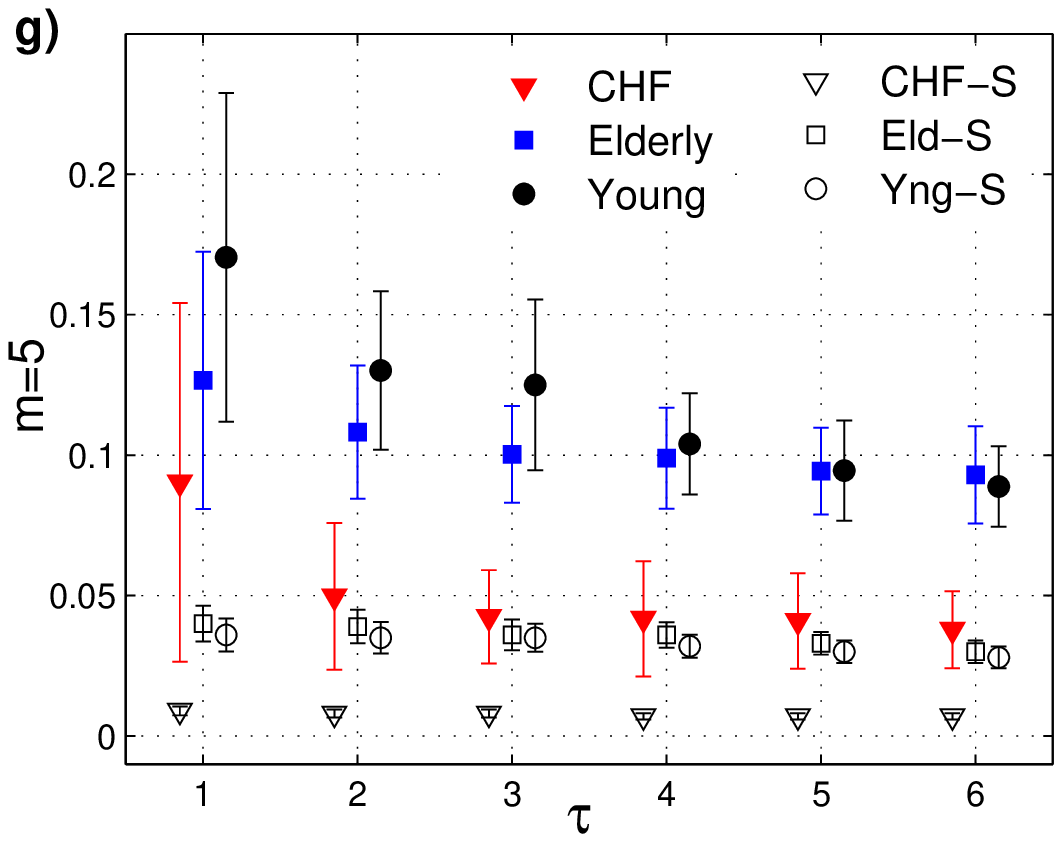}
    \includegraphics[width=4cm,height=4cm]{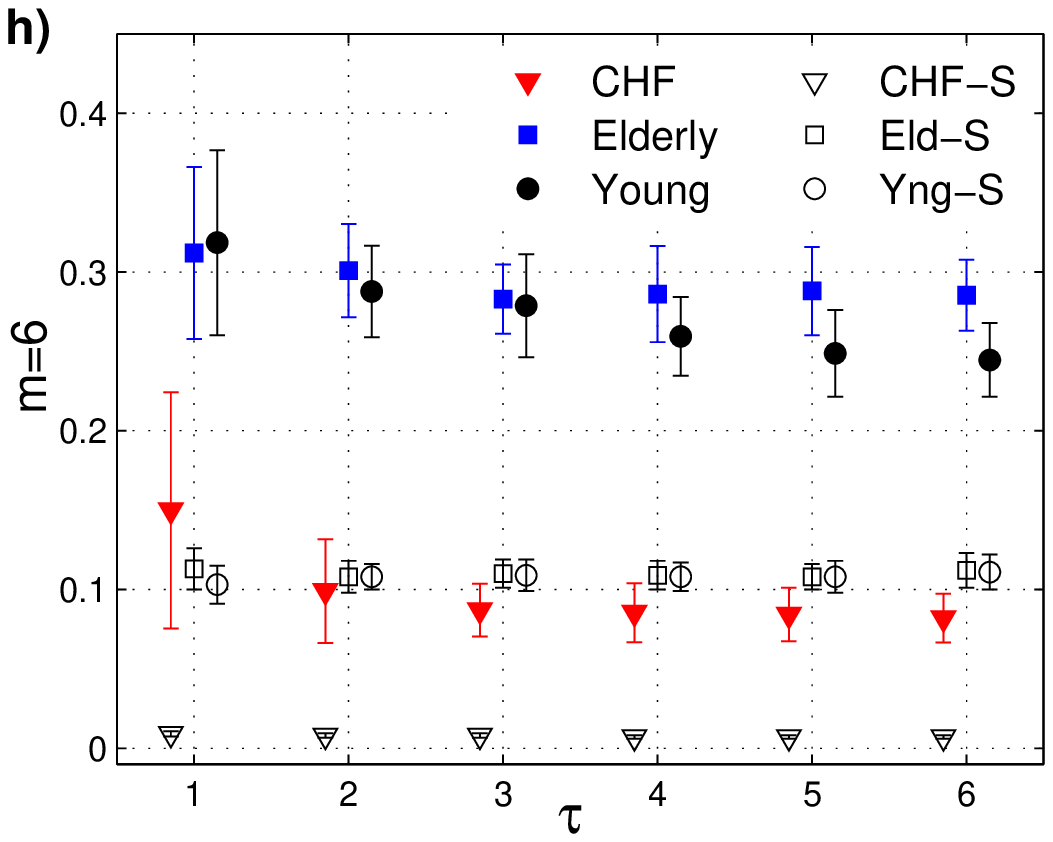}
  \caption{Time irreversibility (mean$\pm$std) of heartbeats and surrogate when $m$=3 to 6 and $\tau$=1 to 6. a), b), c) and d) $Y_{s}$ of original permutations. e), f), g) and h) $Y_{s}$ of equality-based permutations.}
  \label{fig7}
\end{figure}

In this subsection, we test the effect of equal states on the simplified time irreversibility of $m>$2. $Y_{s}$ of probabilistic difference of symmetric permutation of dimension from 3 to 6 and delay from 1 to 6 for the three groups of heartbeats are displayed in Fig.~\ref{fig7}. To ensure the existence of all possible permutations and have reliable outcomes, we recommend the data length to be no less than 8*m!.

Same to Costa index and $Y_{s}$ of $m$=2, $Y_{s}$ of each set of heartbeat is bigger than 97.5th percentile of its 100 sets of surrogate data, and $Y_{s}$ (mean$\pm$std) of all the surrogate data are provided for statistical convenience in Fig.~\ref{fig7}. When $m$=3 and 4, the results share the conclusions of $m$=2. Time irreversibility without considering equal states have contradictory outcomes to the complexity losing theory while that involving equal values have reliable nonlinearity detection whose statistical tests are listed in Tab.~\ref{tab4}.

\begin{table}[htb]
\centering
\caption{Independent sample t test of $Y_{s}$ of equality-based permutations in heartbeats.}
\label{tab4}
\begin{tabular}{cccc|c cc|ccc}
\hline
\multirow{2}{*}{$\tau$}&
\multicolumn{3}{c}{$m$=3}&\multicolumn{3}{c}{$m$=4}&\multicolumn{3}{c}{$m$=5} \\
\cline{2-10}
   &C-E	     &E-Y	    &C-Y     &C-E	    &E-Y	    &C-Y     &C-E	    &E-Y	    &C-Y\\
\hline
2 &0.16645	&0.04805	&1.10E-5  &0.00058	&0.00562	&1.35E-9 &9.17E-12	&0.01354	&6.52E-16\\
3 &8.31E-5	&5.80E-5	&8.83E-8 &5.74E-9	&0.00030	&3.96E-8 &2.15E-18	&0.00303	&2.74E-11\\
4 &0.01759	&0.06108	&9.39E-8 &3.00E-6	&0.01405	&1.25E-11 &2.79E-15	&0.37826	&9.54E-17\\
5 &0.01072	&0.23631	&7.30E-5 &4.95E-6	&0.20714	&2.74E-8 &3.13E-17	&0.87201	&1.75E-16\\
\hline
\end{tabular}
\end{table}

As for the equal-involved $Y_{s}$, when $m$=3 and 4, the best discrimination of the heartbeats lie in $\tau$=3 or 4, and when the delay becomes bigger, the discriminations deteriorate, which is in line with the case of $m$=2. When $m$=5, statistical discriminations between the CHF and the two groups of healthy heartbeats become better, while those between the healthy young and elderly deteriorate when $\tau$ is bigger than 4, the healthy elderly even have higher time irreversibility than the healthy young, which is shared by cases of $m$=6 and $\tau$$>$1. According to our findings and for computational convenience, we would like to recommend $m$ no bigger than 4 for the time irreversibility in heartbeats.

$Y_{s}$ of the probabilities of permutations considering equal values reliably characterize the time irreversibility of heartbeats, while that involving no equality yields results completely contradictory with the conventional wisdom. Equal state in heartbeats should not be neglected not only due to its association with cardiac regulation but also because of its significant effects on the nonlinear dynamics analysis, particularly those based on relative values like the permutation.

\section{Discussions}
Theoretically speaking, there should not be exact equal heartbeat interval if the resolution is high enough, however, there might be equal values in practical applications under low precision of signals collection and R wave detection. Due to the limited time and ADC resolutions, there are equal heartbeats intervals (3.62$\pm$1.80 and 8.05$\pm$4.41 percents in the healthy young and elderly data sets), and under some pathological conditions, like the congestive heart failure, decreased HRV will bring more equal states (20.31$\pm$7.34 percent in the CHF heartbeats). The high distribution of equal RR intervals, brought by the limited time and ADC resolutions, also has close connection with cardiac conditions. Noting that there are other different cardiac intervals (PR, QRS, ST, etc) that carry a lot of important information about cardiac conditions. Our idea of employing the distribution of equal values in RR interval may also expand to measuring changes in the length of these interval, which is worth more detailed and comprehensive research.

To apply the rate of equal RR intervals clinically, it is important to investigate its quality in case of frightening a healthy subject or missing a severe situation. Taking $\tau$=1 as example, if we treat '$eR\leq0.054$' as healthy young, '$eR\geq0.124$' as CHF and '$0.054<eR<0.124$' as healthy elderly, two types of errors of the three kinds of heartbeats are listed in Tab.~\ref{tab5}. In the table, the CHF-Young type I error (false positive) is defined as the probability of CHF being mistreated as healthy young, and the CHF-Young type II error (false negative) implies the probability of healthy young being mistaken as CHF.

\begin{table}[htb]
\centering
\caption{Type I and type II errors of $eR$ in the three groups of heartbeats.}
\label{tab5}
\begin{tabular}{cccc}
\hline
 &CHF-Young	&CHF-Elderly	 &Elderly-Young \\
\hline
Type I &0.000	 &0.114	 &0.250 \\
Type II &0.000	 &0.150	 &0.100 \\
\hline
\end{tabular}
\end{table}

From Tab.~\ref{tab5}, there is no mistreatment between the CHF and healthy young while 5 CHF patients are mistreated as the healthy elderly and 3 healthy elderly people are taken care as CHF patients. Also, 5 healthy elderly volunteers are regarded as the young and 2 healthy young subjects are labeled to be elderly. Therefore the healthy elderly subjects are more likely to be mistaken. Due to the limitation of data sets, we would like to emphasize that $eR$ have to be validated by more representative number of heartbeats.

\begin{figure}[htb]
  \centering
    \includegraphics[width=5.9cm,height=4.5cm]{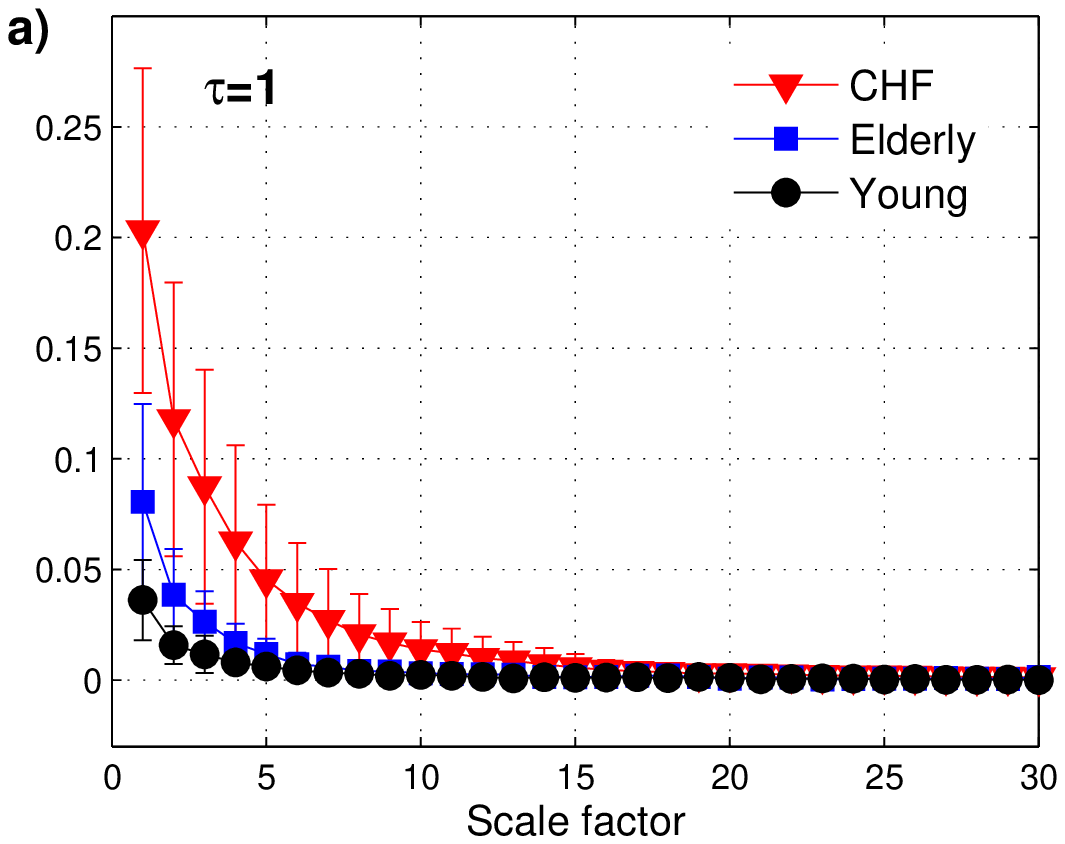}
    \includegraphics[width=5.9cm,height=4.5cm]{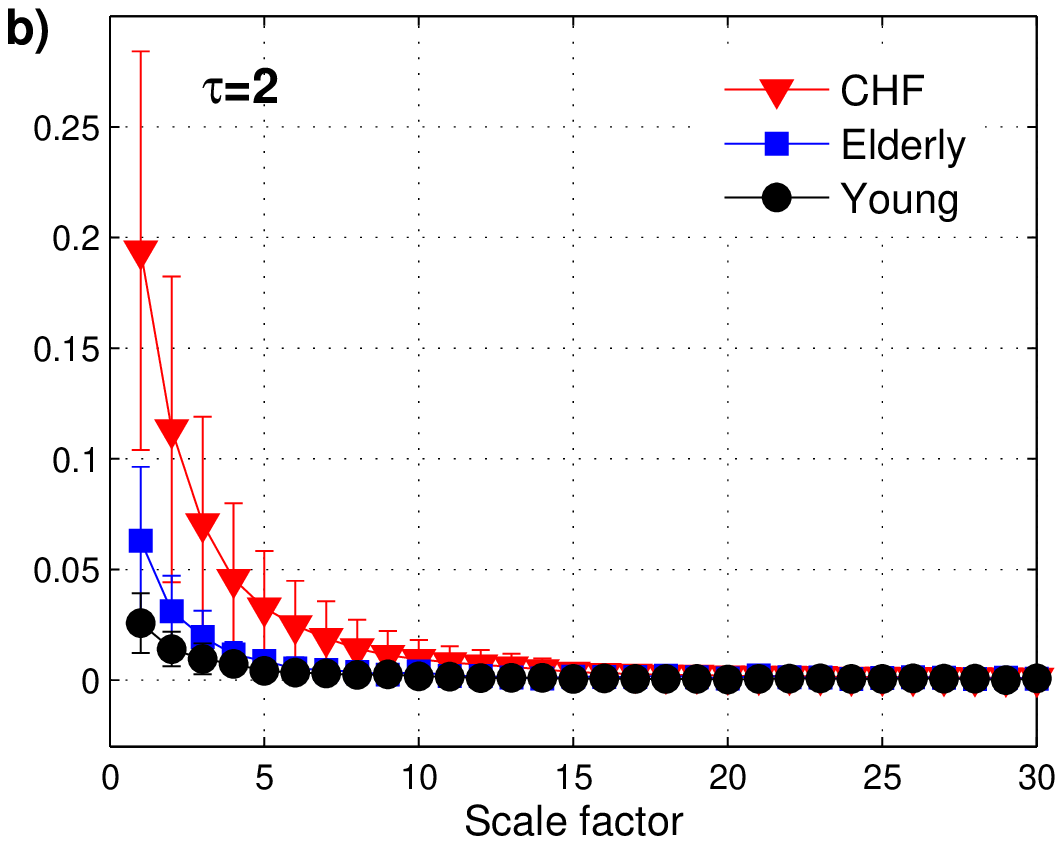}
  \caption{The rates of equal values (mean$\pm$std) of CHF, healthy young and elderly heartbeats . a) $\tau$=1. b) $\tau$=2.}
  \label{fig8}
\end{figure}

Concerning the important information about cardiac autonomic function conveyed by the equal heartbeat intervals, some preprocessing methods, like the multi-scale technique \cite{Costa2002,Costa2008}, for heart rates should be reconsidered. As for the contradictory time irreversibility of the three groups of heartbeats considering no equal states, an explanation might be the multi-scale theory that the single scale method fails to account for the multiple time scales inherent in the healthy systems and lead to paradox. The multi-scale process, constructing scaled $\{y_{j}\}$ as $y_{j}=\frac{1}{s}\Sigma^{js}_{i=(j-1)S +1}x_{i}$ where $s$ is scale factor, is a coarse-graining procedure and has impact on the distribution of equal heartbeat intervals, illustrated in Fig.~\ref{fig8}.

The multi-scale process significantly eliminates the equal states in heartbeats. There is no equal states in the healthy young heartbeats when scale is bigger 30. Of $\tau$=1, when scale is bigger than 7, the healthy young and elderly cannot be discriminated (p$>$0.05), and when scale is bigger than 17 and 18, the discriminations of CHF and healthy young and elderly are not acceptable (p$>$0.05) statistically. Of $\tau$=2, when the scale is larger than 5, the healthy young-elderly differences is not acceptable (p$>$0.05) in statistics. The equal states as well as the physiological and pathological information are eliminated by the coarse-graining multiscale procedure. Therefore, the side effects of multi-scale technique on heartbeats analysis should be paid more attention.

The selection of $m$ and $\tau$ is crucial for the quality of nonlinearity extraction \cite{Casdagli1991} and has significant impact on the permutation-based time irreversibility analysis \cite{Yao2018PLA}. Many scholars have proposed methods to estimate the interrelated $m$ and $\tau$ or alternative parameter like time window length \cite{Kugiumtzis1996,Kim1999}. However, there is no strict standard for the most appropriate delay, some employ ¡®trial and error¡¯ or make choices empirically. According to our analysis in these heartbeats, when $\tau$ is 1, the nonlinear information might be still compressed along the identity line, which is called redundance, and when $\tau$ is 4 or bigger, heartbeats¡¯ nonlinear dynamics are causally disconnected, which is called irrelevance. The most appropriate choice $\tau$=2 or 3 might also represent some relevant frequency in the dynamic of the heart rate, which however need to be validated by more related researches. As for the dimension $m$, it should not be smaller than the inherent dimension of a process from the physical point of view. According to C. Bandt and F. Shiha \cite{2007Bandt}, when $m\geq$5, there are no close formulas for arbitrary order patterns, and order patterns do not fit very well with autocorrelation or spectrum even for Brownian motion and simplest moving-average processes, which is shared by our findings. When $m\geq$5, time irreversibility of the healthy young and elderly heartbeats change and become different as $\tau$ increases. We recommend the dimension of no bigger than 4 and the delay of 2 or 3 in time irreversibility analysis of heartbeats, while the parameters should be investigated and adjusted accordingly in other situations.

\section{Conclusions}
To conclude, there is a significant number of equal values in the discrete heartbeats, and the rate of equal RR interval contains important information about cardiac regulation mechanism and plays a crucial role in the permutation-based time irreversibility analysis.

The distribution of equal heartbeats interval is a simple and feasible indicator for cardiac conditions and contributes to develop relevant biomarkers in the area of heart analysis. The CHF heart rates have significantly higher distribution of equal RR interval than the healthy subjects and the healthy young have lower rate of equal states than the elderly, and the discriminations among the three kinds of heart rates are acceptable even at very short data length. HRV has close connections with cardiac physiological conditions and autonomic regulation, and the reduced HRV, serving as an independent predictor of mortality, could be characterized by the increased distribution of equal heartbeat intervals clinically.

In the heartbeats¡¯ nonlinear dynamics analysis using probabilistic difference between order patterns, neglected equalities lead false conclusions while $Y_{s}$ of permutation considering equal values shows promising nonlinearity detection. And the time irreversibility based on equality-involved permutation further validates the complexity losing theory about diseased and aging heartbeats.

\section{Acknowledgments}
The project is supported by the National Natural Science Foundation of China (Grant Nos. 31671006, 61771251), Jiangsu Provincial Key R\&D Program (Social Development) (Grant No.BE2015700, BE2016773), Natural Science Research Major Program in Universities of Jiangsu Province (Grant No.16KJA310002), Postgraduate Research \& Practice Innovation Program of Jiangsu Province (KYCX17-0788).

\section*{References}

\bibliography{mybibfile}

\end{document}